%% file: ratios-rhic.tex
\documentclass[superscriptaddress,preprint,secnumarabic,amssymb,nobibnotes,aps,prl]{revtex4}
\usepackage{graphicx}
\usepackage{bm}
\input{./define}

\begin{document}
\date{\today}
\title{Particle ratios at RHIC: \\Effective hadron masses and chemical 
freeze-out}
\author{D. Zschiesche} 
\affiliation{Institut f\"ur Theoretische Physik,
        Robert Mayer Str. 8-10, D-60054 Frankfurt am Main, Germany}

\author{S. Schramm}
\affiliation{Argonne National
Laboratory, 9700 S. Cass Avenue, Argonne IL 60439, USA}

\author{J. Schaffner-Bielich}
\affiliation{Department of Physics, Columbia University, 
New York, NY 10027, USA}

\author{H.~St\"ocker}
\affiliation{Institut f\"ur Theoretische Physik,
        Robert Mayer Str. 8-10, D-60054 Frankfurt am Main, Germany}
\affiliation{Laboratoire de Physique Subatomique et des Technologies Associ\'ees \\
University of Nantes - IN2P3/CNRS - Ecole des Mines de Nantes \\
4 rue Alfred Kastler, F-44072 Nantes Cedex 03, France}

\author{W.~Greiner}
\affiliation{Institut f\"ur Theoretische Physik,
        Robert Mayer Str. 8-10, D-60054 Frankfurt am Main, Germany}

\begin{abstract}
The measured particle ratios in central heavy-ion collisions at 
RHIC-BNL are investigated within a chemical and 
thermal equilibrium  
chiral $SU(3) \; \sigma-\omega$ approach. The commonly adopted 
noninteracting gas calculations yield temperatures close to or
above the critical temperature 
for the chiral phase transition, but without taking into account any
interactions.
Contrary, the chiral $SU(3)$ model predicts temperature and density
dependent effective hadron masses 
and effective chemical potentials in the medium and a transition to a chirally
restored phase at high temperatures or chemical potentials. 
Three different parametrizations of the 
model, which show different types of phase transition behaviour, 
are investigated.
We show that if a chiral phase transition occured in those collisions,
''freezing'' of the relative hadron abundances in the symmetric
phase is excluded by the data. Therefore, either very rapid chemical
equilibration must occur in the broken phase, or the measured hadron
ratios are the outcome of the dynamical symmetry breaking.
Furthermore, the extracted chemical freeze-out parameters  
differ considerably from those obtained in simple noninteracting gas
calculations. In particular, the three models yield up to 
$35 \mbox{ MeV}$ lower temperatures than the free gas
approximation. The in-medium masses turn out differ up to $150$~MeV
from their vacuum values.
\end{abstract}
\draft
\maketitle
\section{Introduction}
Thermodynamical equilibrium 
calculations of particle production in high energy  
particle- and nuclear collisions 
have been carried out for a long time 
\cite{ferm50,land53,west76,stoe78,cser86,hahn86,brau96,Let00,raf01,bec00,brau01}.  
Recently hadron abundances and particle ratios have been measured in 
heavy-ion collisions from 
SIS, AGS, SPS to RHIC energies.
These data have revived the interest in the extraction of temperatures
and chemical potentials from thermal equilibrium ''chemical'' model 
analyses. The experimentally determined hadron ratios can be fitted 
well with straightforward noninteracting gas model calculations 
\cite{brau96,Let00,brau98,cley99,raf01,bec00,brau01}, if a
sudden breakup of a thermalized source is assumed and once the
subsequent feeding of the various channels by the strongly decaying
resonances is taken into account. From the 
$\chi^2$ freeze-out fits one has constructed a quite narrow band of 
freeze-out values in the $T-\mu_B$ plane (see e.g.\cite{brau98,cley99}).
The extracted freeze-out
parameters are fairly close to the phase transition curve for SPS and RHIC energies. 
However, when we are indeed so close to the phase transition or to a
crossover as suggested by the \cal{data} for $T$ and $\mu_B$, we can
not afford to neglect the very in-medium effects we are after - and
which, after all, do produce the phase transition. Thus, 
since noninteracting gas models neglect any kind of possible in-medium
modifications they can not yield information about the phase transition.

Therefore, we will employ below a relativistic selfconsistent
chiral model of hadrons and hadron matter developed in 
\cite{paper2,paper3, springer}. 
This model can be used as a thermodynamically consistent 
effective theory or as a toy model, which embodies the restoration of
chiral symmetry at high temperatures or densities. Therefore the model
predicts temperature
and density dependent hadronic masses and effective chemical
potentials, which have already been proposed and considered in 
\cite{stoe78,thei83,scha91,springer,BroRho,mich01}. 
Thus, using the chiral $SU(3)$ model we can investigate, 
whether the freeze-out in fact takes place close to
the phase transition boundary (if it exists) and if the extracted 
$T, \mu_B$ parameters are strongly model dependent.
Depending on the chosen parameters and
degrees of freedom different scenarios for the chiral phase change are
predicted by the model: Strong or weak first order phase transition or
a crossover. The transitions take place around 
$T_c = 155 \mbox{ MeV} $ \cite{zsch01,springer}, which is in qualitative
agreement with lattice predictions
\cite{Karsch:2000kv} for the critical temperature for the onset of a 
deconfined phase which coincides with that of a chirally restored
phase \cite{kar98}. 

\section{Model description}
\begin{figure}
\begin{center}
\centerline{\parbox[b]{8cm}{
\includegraphics[width=9.2cm,height=9cm]{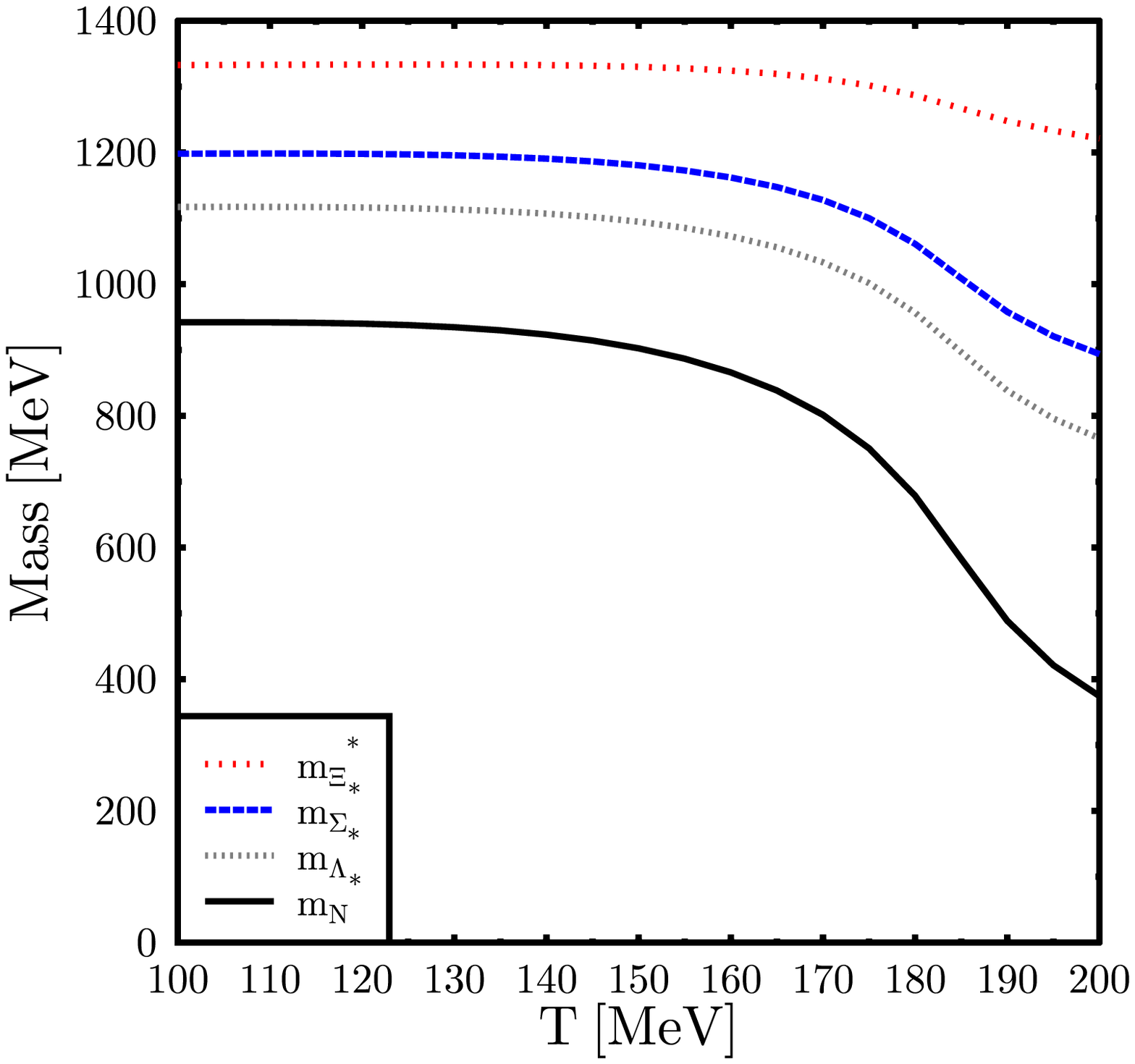}}
\parbox[b]{8cm}{
\includegraphics[width=9.2cm,height=9cm]{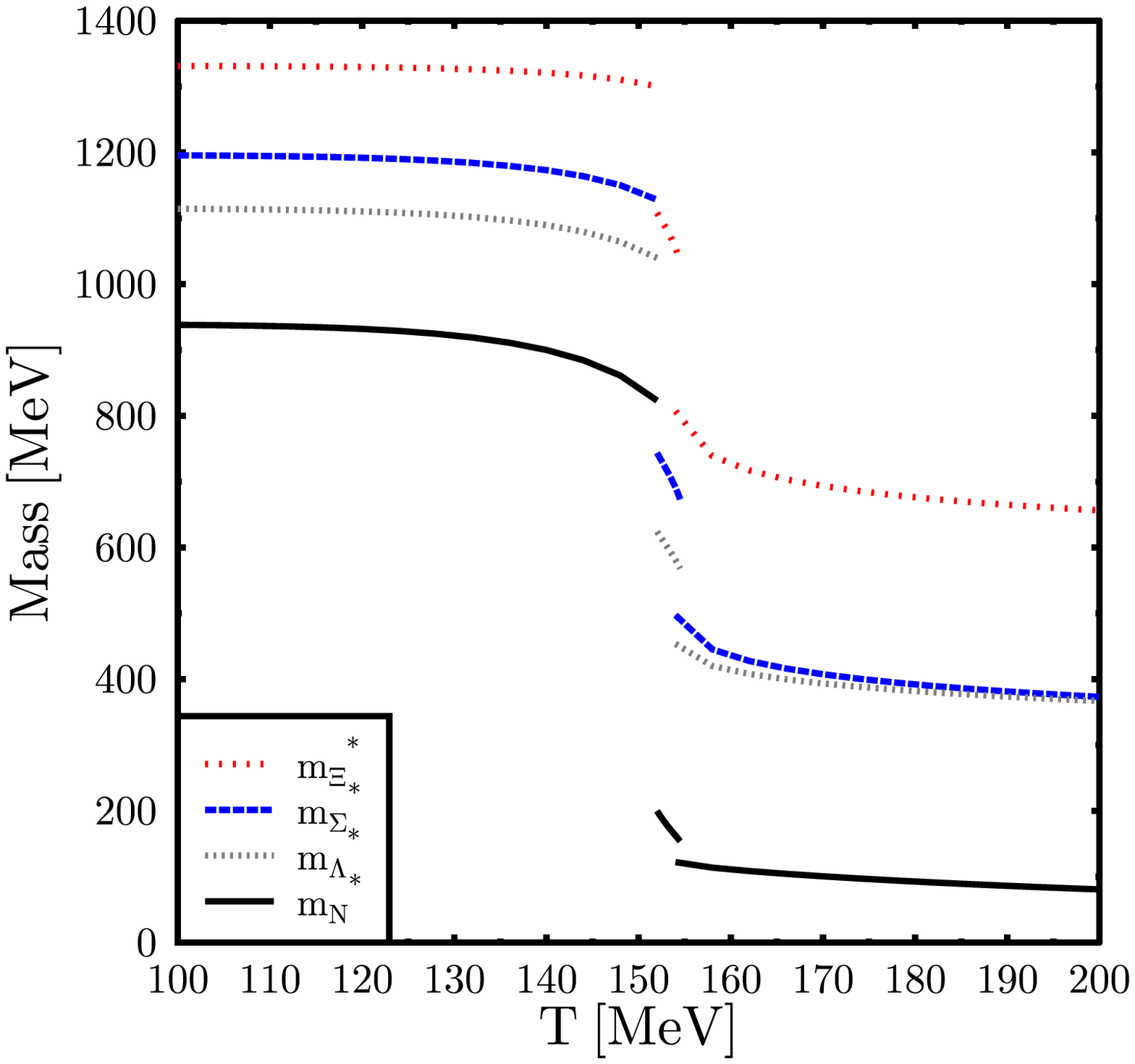}}}
\vspace{-1cm}
\caption{\label{bmasses}Baryon octet masses as function of temperature for vanishing
chemical potential. Left CI, right CIII. 
Note the continuous change of the masses starting around  
$T = 150 \mbox{ MeV}$.
In contrast for CIII two phase-transitions occur around 
$T_c \approx 155 \mbox{ MeV}$.
These result from the separate jumps in the nonstrange ($\sigma$) and
the strange ($\zeta$) condensate.}
\end{center}
\end{figure}

The chiral $SU(3)$ model is presented in detail in 
\cite{paper3, springer}. We will briefly introduce the
model here:
We consider a relativistic field theoretical model of 
baryons and mesons built on
chiral symmetry and broken scale invariance. The general form of the
Lagrangean looks as follows:
\be
\label{lagrange}
{\cal L} = {\cal L}_{\mathrm{kin}}+\sum_{W=X,Y,V,{\cal A},u}{\cal L}_{\mathrm{BW}}+
{\cal L}_{\mathrm{VP}}
+{\cal L}_{\mathrm{vec}}+{\cal L}_{0}+{\cal L}_{\mathrm{SB}} .\no
\ee
${\cal L}_{\mathrm{kin}}$ is 
the kinetic energy term, ${\cal L}_{\mathrm{BW}}$ includes the  
interaction terms of the different baryons with the various spin-0 and spin-1 
mesons (see \cite{paper3} for details). 
The baryon masses are generated by both, the nonstrange
$\sigma$ (${<q\bar{q}>}$)  
 and the strange $\zeta$ (${<s\bar{s}>}$) scalar condensate. 
${\cal L}_{\rm{VP}}$ contains the interaction terms 
of vector mesons with pseudoscalar mesons. 
${\cal L}_{\rm{vec}}$ generates the masses of the spin-1 mesons through 
interactions with spin-0 fields, and ${\cal L}_{0}$ gives the meson-meson 
interaction terms which induce the spontaneous breaking of chiral symmetry.
It also includes a scale-invariance breaking logarithmic potential. Finally, 
${\cal L}_{\mathrm{SB}}$ introduces an explicit symmetry breaking of the
U(1)$_A$, the SU(3)$_V$, and the chiral symmetry. 
All these terms have been discussed in detail in \cite{springer,paper3}.\\
The hadronic matter
properties at finite density and temperature are studied in 
the mean-field approximation \cite{serot97}. 
Then the Lagrangean (\ref{lagrange}) 
becomes
\begin{eqnarray}
{\cal L}_{BX}+{\cal L}_{BV} &=& -\sum_{i} \overline{\psi_{i}}[g_{i 
\omega}\gamma_0 \omega^0 
+g_{i \phi}\gamma_0 \phi^0 +m_i^{\ast} ]\psi_{i} \\ \no
{\cal L}_{vec} &=& \frac{ 1 }{ 2 } m_{\omega}^{2}\frac{\chi^2}{\chi_0^2}\omega^
2  
 + \frac{ 1 }{ 2 }  m_{\phi}^{2}\frac{\chi^2}{\chi_0^2} \phi^2
+ g_4^4 (\omega^4 + 2 \phi^4)\\
{\cal V}_0 &=& \frac{ 1 }{ 2 } k_0 \chi^2 (\sigma^2+\zeta^2) 
- k_1 (\sigma^2+\zeta^2)^2 
     - k_2 ( \frac{ \sigma^4}{ 2 } + \zeta^4) 
     - k_3 \chi \sigma^2 \zeta \\ 
&+& k_4 \chi^4 + \frac{1}{4}\chi^4 \ln \frac{ \chi^4 }{ \chi_0^4}
 -\frac{\delta}{3} \chi^4 \ln \frac{\sigma^2\zeta}{\sigma_0^2 \zeta_0} \\ \no
{\cal V}_{SB} &=& \left(\frac{\chi}{\chi_0}\right)^{2}\left[m_{\pi}^2 f_{\pi} 
\sigma 
+ (\sqrt{2}m_K^2 f_K - \frac{ 1 }{ \sqrt{2} } m_{\pi}^2 f_{\pi})\zeta 
\right] , 
\end{eqnarray}
where $m_i$ is the effective mass of the hadron species $i$.  
$\sigma$
and $\zeta$ correspond to  
the scalar condensates, $\omega$ and $\phi$ represent
the non-strange and the strange vector field respectively, and $\chi$ is
the scalar-isoscalar dilaton field, which mimics the
 effects of the gluon condensate \cite{sche80}. Only 
the scalar (${\cal L}_{BX}$) and the vector meson terms (${\cal L}_{BV}$)
contribute to the baryon-meson interaction, since for all other mesons
the expectation value vanishes in the mean-field approximation.
The grand canonical potential  
$\Omega$ per volume $V$ 
as a function of chemical potential $\mu$ and temperature $T$ can 
be written as:
\begin{equation}
   \frac{\Omega}{V}= -{\cal L}_{vec} - {\cal L}_0 - {\cal L}_{SB}
-{\cal V}_{vac} \\
\mp  T \sum_i \frac{\gamma_i }{(2 \pi)^3}
\int d^3k \left[\ln{\left(1 \pm e^{-\frac 1T[E^{\ast}_i(k)-\mu^{\ast}_i]}\right)}
\right], 
\end{equation}
with the baryons (top sign) and mesons (bottom sign).
The vacuum energy ${\cal V}_{vac}$ (the potential at $\rho_B=0, T=0$) 
has been subtracted in 
order to get a vanishing vacuum energy. $\gamma_i$ denote the 
hadronic spin-isospin degeneracy factors.
The single particle energies are 
$E^{\ast}_i (k) = \sqrt{ k_i^2+{m_i^*}^2}$ 
and the effective chemical potentials read
 $\mu^{\ast}_i = \mu_i-g_{i \omega} \omega-g_{\phi i} \phi$.\\ 
%
The mesonic fields are determined by extremizing $\frac{\Omega}{V}(\mu, T=0)$.
The density of particle $i$ can be calculated by differentiating
$\Omega$ with respect to the corresponding chemical potential $\mu_i$.
This yields:
\begin{eqnarray}
\rho_i = \gamma_i
\int \frac{d^3 k}{(2 \pi)^3}
\left[
\frac 1{\exp{[(E_i^\ast-\mu_i^\ast)/T]\pm1}}
\right] .
\end{eqnarray}
All other thermodynamic quantities can also be obtained from the grand
canonical potential.
In the present calculation the lowest lying baryonic octet and
decuplet and the 
lowest lying mesonic nonets
are coupled to the relativistic mean 
fields. Depending on the coupling of the baryon resonances (the
decuplet) to the
field equations, the 
model shows a first order phase transition or a crossover (for details see \cite{zsch01}).
We will use three different parameter sets: Parameter set CI treats the members
of the baryon decuplet as free particles, which yields a
crossover behaviour. Parameter sets CII and CIII include also 
the (anti)-baryon decuplet as sources for the meson field
equations. They differ by an additional explicit
symmetry breaking for the baryon resonances along the hypercharge direction,  
as described in \cite{paper3} for the baryon octet. This is included in
CII and not used 
in CIII. This leads to a weak first order phase transition at $\mu=0$
for CII and two first order phase transitions for CIII, which can be
viewed as one strong first order phase transition.
Heavier resonances up to $m=2$~GeV are always included
as free particles. 
The resulting baryon masses for CI and CIII are shown in fig. \ref{bmasses}.
We observe a continous decrease of the baryon masses for 
CI starting at  $T \approx 150 \mbox{ MeV}$. 
In contrast, CIII shows two jumps  
around $T = 155 \mbox{ MeV}$.
The critical energy densities, the entropy densities and the transition
temperatures for $\mu_q=\mu_s=0$ 
($\mu_q=\mu_B/3, \mu_s=\mu_B/3-\mu_S  $)
 are specified in table \ref{enentdensities}.
\begin{table}[h]
\begin{center}
\bt{c|c|c|c|c|c}
     & ${\epsilon^-/\epsilon_0}$ & ${\epsilon^+/\epsilon_0}$  
     & $s^- [fm^{-3}]$ & $s^+ [fm^{-3}]$ 
     & $T_c [\rm{MeV}]$ \\
\hline
CII           & 2.8  & 7.2  & 2.8  & 6.7 & 156.3    \\ 
CIII - 1st PT & 2.3  & 8.3  & 2.4  & 7.9  & 153.4    \\
CIII - 2nd PT & 10.5 & 17.1  & 9.8  & 15.7 & 155.5  \\
 \et
\caption{\label{enentdensities} \textrm{ 
Energy density, entropy density and phase transition temperatures  
for CII,CIII, $\mu_q = \mu_s = 0$.
The $(-),(+)$ signs refer to an approach to the phase
transition from  below and above, respectively. $T_c$ denotes the phase transition
temperature.
$\epsilon_0 = 138.45$~MeV/fm$^3$ denotes the energy density of nuclear
matter in the ground state.}}
\end{center}
\end{table}

\section{Particle ratios in the chiral $SU(3) \times SU(3)$ model}
Since the chiral SU(3) model 
predicts density and temperature dependent hadronic masses and
effective potentials, in contrast to noninteracting models, 
the resulting particle ratios and therefore 
the deduced freeze-out 
temperatures and baryon chemical potentials are
expected to change \cite{zsch00}. Hence 
in the following, we identify combinations of temperatures and
chemical potentials that fit the observed particle ratios in the
chiral model. In all calculations the value of the strange chemical
potential $\mu_S$ is chosen such that the net strangeness $f_s=0$.
We are looking for minima of $\chi^2$ with
\begin{equation}
\chi^2 = \sum_i \frac {\left(r_i^{exp} - r_i^{model}\right)^2}{\sigma_i^2}.
\end{equation}
Here $r_i^{exp}$ is the experimental ratio, $r_i^{model}$ is the ratio
calculated in the model and $\sigma_i$ represents the error in the
experimental data points.
We use the same ratios as in 
\cite{brau01}: $\bar{p}/p$, $\bar{\Lambda}/\Lambda$, 
$\bar{\Xi}/{\Xi}$, ${\pi^-}/{\pi^+}$, ${K^-}/{K^+}$,
${K^-}/{\pi^-}$, ${K_0^\ast}/{h^-}$, ${\bar{K_0^\ast}}/{h^-}$.

Even though the only parameters in a thermal and
chemical equilibrium approach on first sight
are the temperature and the baryon chemical potential, 
there exist further unknowns: 
On the one hand, some decays of high mass resonances are
not well known and on the other hand the effect of weak decays in the  
experiments   
strongly depends on the detector geometry and on the reconstruction
efficiency of the experiments. The feeding correction  
from the strong and electromagnetic decays of the hadronic
resonances used here employs 
the procedure used in the UrQMD model \cite{bass98,urqmdhome}.
Weak decays are not considered here. 
We rather focus on the principal question 
whether an interacting chiral
SU(3) approach with $m^{\ast} \neq m_{vac}$ 
can at all  describe the particle yields at RHIC.
Fine tuning of the $\chi^2$ by adjustment of the 
weak decay scheme is not our intention. Even though it has 
been shown \cite{mich_dr} that $\chi^2$ values 
may be improved by including weak decays.

To compare the quality of the fits
obtained in the chiral model with those obtained from the noninteracting
gas approach, we set all masses and chemical potentials 
contained in the chiral model to their 
vacuum values and again use the same UrQMD feeding procedure as for the
interacting model. This yields the ideal gas denoted
$ig_{FFM}$. We find that the resulting ideal gas ratios 
are not identical but comparable to those obtained in the literature  
\cite{zsch00,mich_dr,brau01,flor01}. The differences should only result from a 
different treatment of weak interactions and from the uncertainty 
in the decay scheme of high mass resonances. 

\section{Results for Au + Au Collisions at RHIC}
%
%
%
First, we find that  a reasonable fit of the measured particle ratios
at RHIC is possible in all three phase transition scenarios of the
chiral model and the ideal gas case with comparable quality. 

Second, the resulting freeze-out values depend on
the model employed, i.e. crossover, weak first order, strong first
order or free thermal gas. 

Third, a reasonable description of the data is impossible
above $T_c$ in the models showing a
first order phase transition. This shows that no direct freeze-out
from the restored phase
is observed. 

\begin{figure}[h]
\vspace{-1cm}
\begin{center}
\centerline{\parbox[b]{8cm}{
\includegraphics[width=9.2cm,height=9cm]{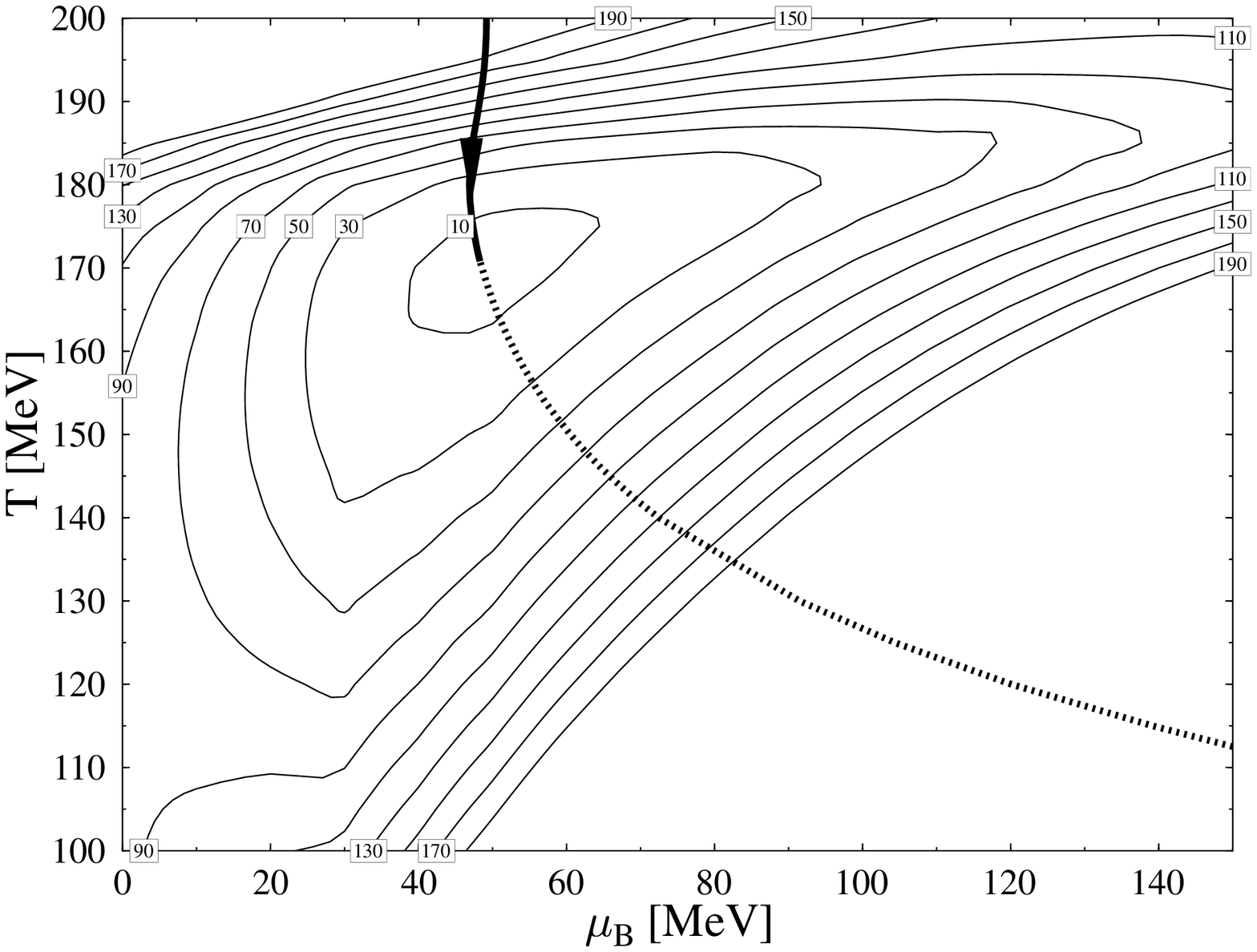}}
\parbox[b]{8cm}{
\includegraphics[width=9.2cm,height=9cm]{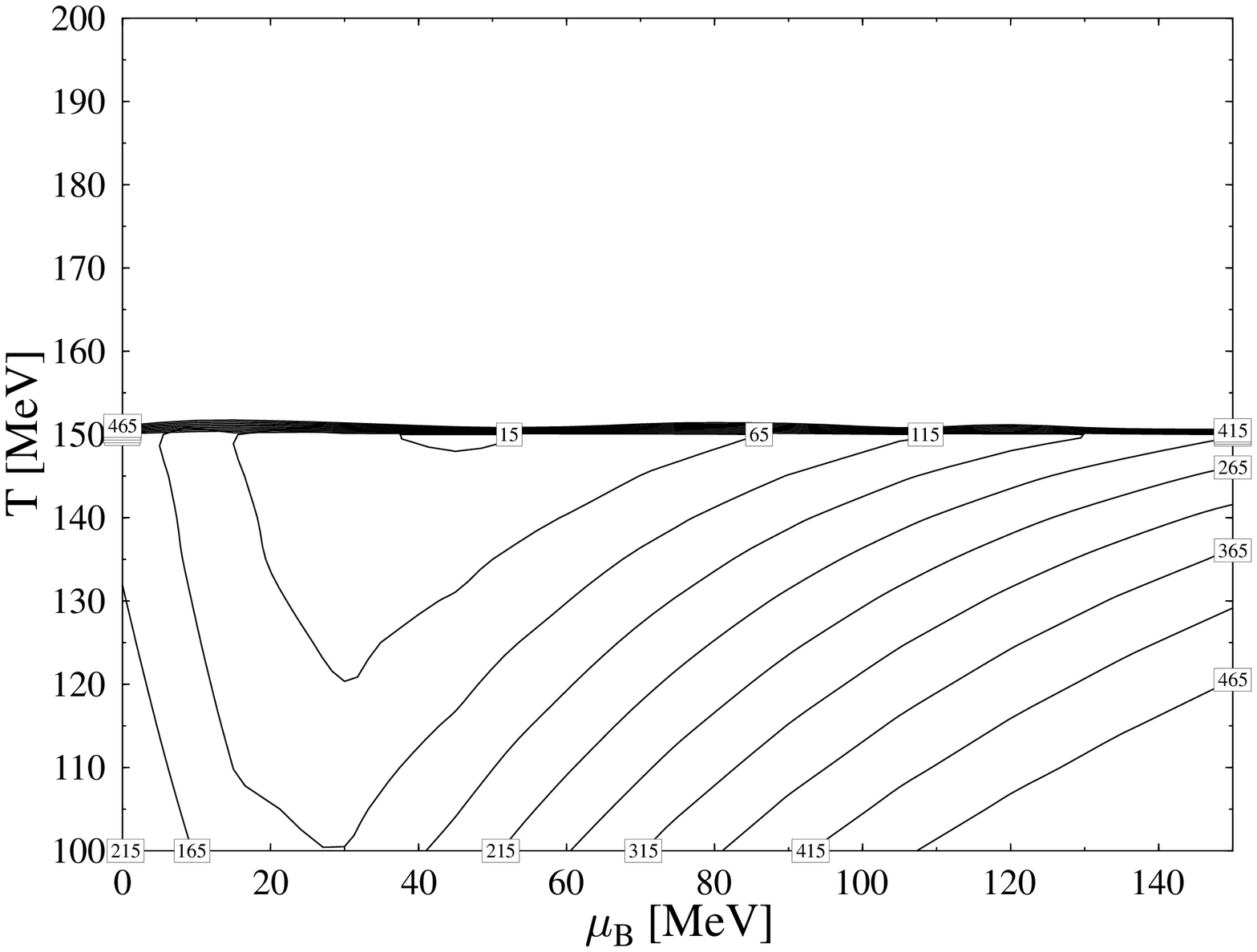}}}
\vspace{-1cm}
\caption{\label{chi2contour}$\chi^2$ contours in the $T-\mu_B$ plane
  for CI (left) and CIII (right).
 Data are taken from \protect\cite{brau01}.
On the left, the adiabatic path (constant entropy
per net baryon $S/A$), corresponding to expanisve cooling of an ideal
fluid, is also shown.
$\mu_S$ is chosen such that $f_s=0$.  }
\end{center}
\vspace{-0.5cm}
\end{figure}

Figure \ref{chi2contour} shows the value of $\chi^2$ in the $T-\mu_B$
plane for the crossover case and for the strong first order phase transition. 
We see that the best fit $T-\mu_B$ values differ in both models. Furthermore, 
in the crossover case $\chi^2$ 
is well behaved as a function of $T$ and $\mu_B$. 
In contrast, the model with a strong first order phase
transition shows a very steep increase of $\chi^2$ at the phase
transition boundary: the quality of the fit decreases drastically
due to the jump of the effective masses
at the phase transition boundary. Above $T_c$ the 
$\chi^2$ values are inacceptable, $\chi^2 > 500$.

\begin{figure}[h]
\vspace*{-1cm}   
\centerline{\parbox[b]{8cm}{
\includegraphics[width=9.2cm,height=9cm]{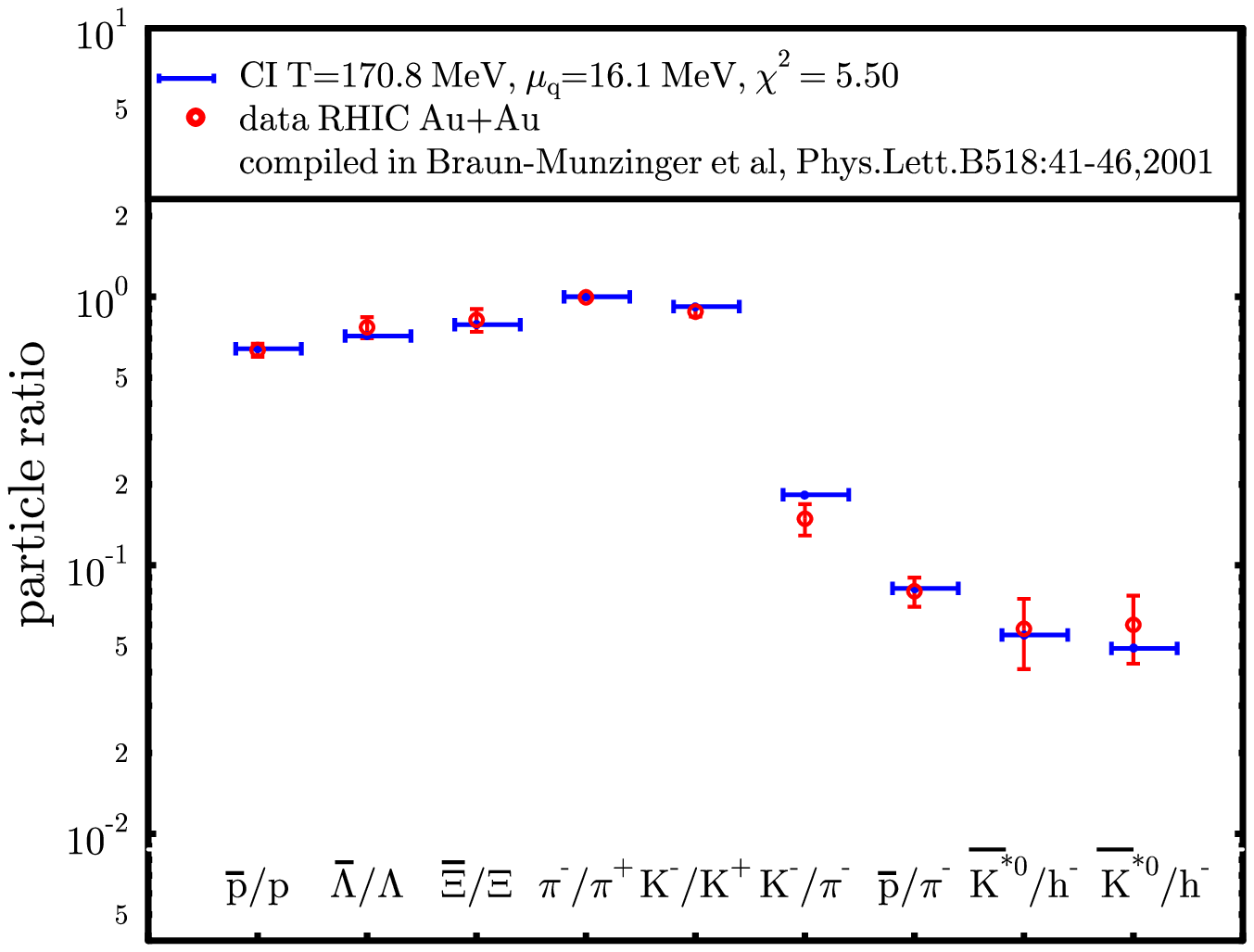}}
\parbox[b]{8cm}{
\includegraphics[width=9.2cm,height=9cm]{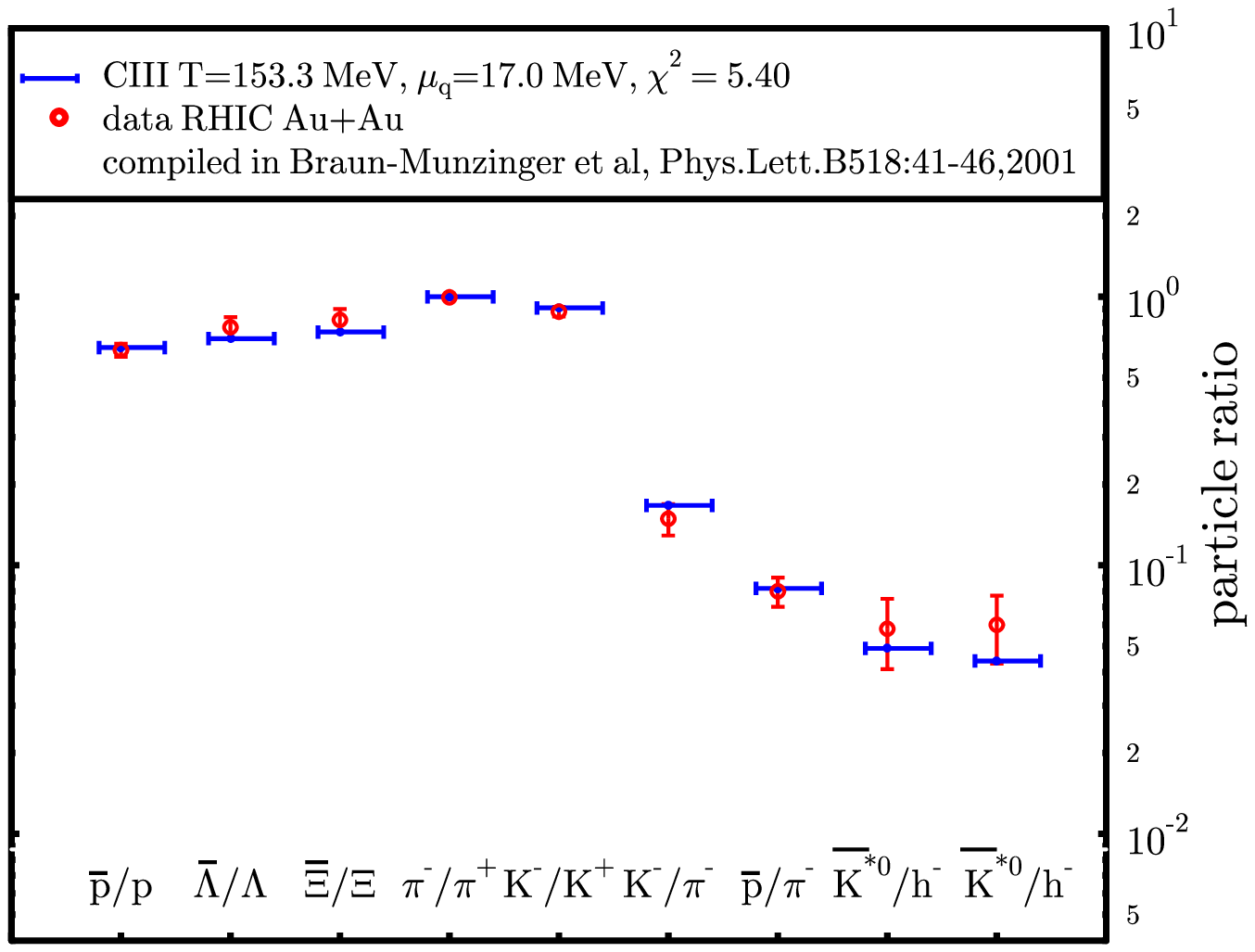}}}
\vspace*{-0.5cm}
\caption{\label{ratios_rhic}
Particle ratios calculated with CI (left) and CIII (right) compared to 
RHIC data as compiled in \protect{\cite{brau01}}.}
\end{figure}

The resulting best-fit particle ratios, $\chi^2$-values and
thermodynamic quantities are shown in 
table \ref{rhic-ratios1} and figure \ref{ratios_rhic}. 

\begin{table}
\begin{center}
\begin{tabular}{|l|l|l|c|c|c|c|}
\hline
{\bf Au+Au} & Experiment & CI & CII & CIII & $ig_{FFM}$ & BMRS \\ \hline \hline
$T_{chem}$ [MeV] & &
\multicolumn{1}{|r|}{{\bf 170.8}} &  
\multicolumn{1}{|r|}{{\bf 155.0}} &
\multicolumn{1}{|r|}{{\bf 153.3}} &
\multicolumn{1}{|r|}{{\bf 187.6}} &
\multicolumn{1}{|r|}{{\bf 174.0}} 
 \\ 
\hline
$\mu _{chem}^{B}$ [MeV] & & 
\multicolumn{1}{|r|}{{\bf 48.3}} &  
\multicolumn{1}{|r|}{{\bf 54.6}} &
\multicolumn{1}{|r|}{{\bf 51.0}} &
\multicolumn{1}{|r|}{{\bf 44.1}} &
\multicolumn{1}{|r|}{{\bf 46.0}}\\ 
\hline 
$\mu _{chem}^{s}$ [MeV] & &
\multicolumn{1}{|r|}{\bf 11.1} &  
\multicolumn{1}{|r|}{\bf 9.8} &
\multicolumn{1}{|r|}{\bf 9.4} &
\multicolumn{1}{|r|}{\bf 13.5} & 
\multicolumn{1}{|r|}{\bf 13.6} \\ 
\hline
$\chi ^{2}$ & & 
\multicolumn{1}{|r|}{{\bf 5.5}} &
\multicolumn{1}{|r|}{{\bf 5.7}} &
\multicolumn{1}{|r|}{{\bf 5.4}} &
\multicolumn{1}{|r|}{{\bf 5.7}} &
\multicolumn{1}{|r|}{{\bf 5.7}} \\ 
\hline 
\hline
$\rho_{had} [fm^{-3}] $ & & 
\multicolumn{1}{|r|}{\bf 0.66} &  
\multicolumn{1}{|r|}{\bf 0.38} &
\multicolumn{1}{|r|}{\bf 0.35} &
\multicolumn{1}{|r|}{\bf 1.12} & 
\multicolumn{1}{|r|}{\bf } \\ 
\hline
$\rho_{B}+\rho_{\bar{B}} \,[fm^{-3}] $ & & 
\multicolumn{1}{|r|}{\bf 0.15} &  
\multicolumn{1}{|r|}{\bf 0.08} &
\multicolumn{1}{|r|}{\bf 0.07} &
\multicolumn{1}{|r|}{\bf 0.28} & 
\multicolumn{1}{|r|}{\bf } \\ 
\hline
$p [MeV/fm^{3}] $ & & 
\multicolumn{1}{|r|}{\bf 108} &  
\multicolumn{1}{|r|}{\bf 55} &
\multicolumn{1}{|r|}{\bf 51} &
\multicolumn{1}{|r|}{\bf 207} & 
\multicolumn{1}{|r|}{\bf } \\ 
\hline
$\epsilon [MeV/fm^{3}] $ & & 
\multicolumn{1}{|r|}{\bf 695} &  
\multicolumn{1}{|r|}{\bf 356} &
\multicolumn{1}{|r|}{\bf 326} &
\multicolumn{1}{|r|}{\bf 1324} & 
\multicolumn{1}{|r|}{\bf } \\ 
\hline
$E/A [MeV] $ & & 
\multicolumn{1}{|r|}{\bf 1053} &  
\multicolumn{1}{|r|}{\bf 937} &
\multicolumn{1}{|r|}{\bf 931} &
\multicolumn{1}{|r|}{\bf 1182} & 
\multicolumn{1}{|r|}{\bf $\approx$ 1100} \\ 
\hline
$S/A $ & &  
\multicolumn{1}{|r|}{\bf 157} &  
\multicolumn{1}{|r|}{\bf 164} &
\multicolumn{1}{|r|}{\bf 177} &
\multicolumn{1}{|r|}{\bf 142} & 
\multicolumn{1}{|r|}{\bf } \\ 
\hline
\hline
$\overline{p}/p$ & 
\begin{tabular}{ll}
$0.65(7)$ [STAR], & $0.64(8)$[PHENIX]  \\
$0.60(7)$ [PHOBOS], &  $0.61(6)$ [BRAHMS]
\end{tabular}
& 0.640 & 0.648 & 0.652 & 0.629 & 0.629\\ \hline
$\overline{\Lambda }/\Lambda $ & $0.77(7)$ [STAR] 
& 0.714 & 0.695 & 0.702 & 0.721 & 0.753 \\ \hline
$\overline{\Xi }/\Xi $ & $0.82(8)$  [STAR] 
& 0.787 & 0.731 & 0.743 & 0.834 & 0.894 \\ \hline
$\pi ^{-}/\pi ^{+}$ & 
\begin{tabular}{ll}
$1.00(2)$ [PHOBOS], & $0.95(6)$[BRAHMS]
\end{tabular} 
& 1.000 & 1.000 & 1.000 & 1.000 & 1.007\\ \hline
$K^{-}/K^{+}$ &
\begin{tabular}{ll}
$0.88(5)$ [STAR], & $0.78(13)$ [PHENIX] \\
$0.91(9)$ [PHOBOS], & $0.89(7)$ [BRAHMS]
\end{tabular}
& 0.919 & 0.914 & 0.915 & 0.916 & 0.894 \\ \hline
$K^{-}/\pi ^{-}$ & $0.15(2)$ [STAR] 
& 0.183 & 0.168 & 0.168 & 0.179 & 0.145\\ \hline 
$\overline{p}/\pi ^{-}$ & $0.08(1)$ [STAR] 
& 0.082 & 0.084 & 0.078 & 0.083 & 0.078 \\ \hline
$\overline{K_{0}^{\ast }}/h^{-}$ & $0.058(17)$ [STAR] 
& 0.055 & 0.049 & 0.049 & 0.046 & 0.032\\ \hline
$K_{0}^{\ast }/h^{-}$ & $0.060(17)$ [STAR] 
& 0.049 & 0.044 & 0.044 & 0.041 & 0.037\\ \hline
\end{tabular}
\end{center}
\caption{{\small Chiral fit of the particle ratios measured at RHIC
at $\sqrt{s}=130$ GeV. Data and BMRS-fit taken from \protect{\cite{brau01}}.}}
\label{rhic-ratios1}
\end{table}
The $\chi^2$ values for the chiral model are: 
$\chi^2_{CI} = 5.50$,  $\chi^2_{CII} =5.73$ and 
$\chi^2_{CIII} = 5.40$. 
Thus, all three parameter sets describe the data equally
well. Furthermore, the agreement is as good as in the noninteracting gas
calculation ($\chi^2_{ig}= 5.72$ \cite{brau01}, $\chi^2_{FFM}=5.66$).  
The best fit $T-\mu_B$ parameters vary quite considerably between the different
models. The noninteracting gas calculation yields $T=187.6 \mbox{
MeV}$ and $\mu_B=44.1 \mbox{ MeV}$. These freeze-out values can be 
compared to those obtained in other ideal gas calculations: 
$T=174\mbox{ MeV}, \mu_B=46 \mbox{ MeV} $ in \cite{brau01}, 
$T=165\mbox{ MeV}, \mu_B=41 \mbox{ MeV} $ in \cite{flor01} and 
$T=190\mbox{ MeV}, \mu_B=45 \mbox{ MeV} $ in \cite{xu01}.
The crossover case in the interacting chiral model (CI) yields
$T=170.8 \mbox{ MeV}, \mu_B=48.3 \mbox{ MeV}$. Very strong deviations
are found for the models with a first order phase transition (CII,CIII): 
The freeze-out temperatures are $T=155$~MeV (CII) and
$T=153.3$~MeV (CIII), more than $30$~MeV lower
than for $ig_{FFM}$. The fitted baryon chemical potentials $\mu_B$ 
increase by about $7-10 \mbox{ MeV}$. 
These $T-\mu_B$ pairs are very close to the phase
boundary (CII) or even right on it (CIII) and are about $10$ MeV higher
than the values obtained at SPS-energies \cite{zsch02}. Mainly due to
the different freeze-out temperatures the values of the corresponding
thermodynamic quantities vary between the different
approaches. However, 
the energy per particle $E/A$ is approximately $1$~GeV in all
cases. This 'unified freeze-out condition' has already been proposed
in \cite{cley98}. 

\begin{figure}[h]
\vspace*{-0.5cm}   
\includegraphics[width=9cm]{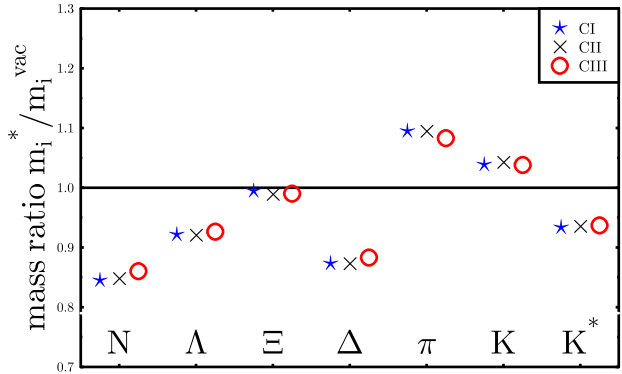}
\vspace*{-0.5cm}
\caption{\label{massen}
Effective masses for the different interacting chiral models and the
ideal gas (vacuum values) case. The differences among the interacting
approaches are less than $2\%$.  }
\end{figure}

The fact that the freeze-out appears right at the phase boundary or at
crossover implies that there are large in-medium corrections, in
particular for the effective masses, a phenomenon observed already in 
\cite{thei83}.
The effective masses shown in figure 
\ref{massen} are shifted up to $15 \%$ from
their vacuum values. However, all the interacting models show similar values
for the effective mass of a given hadron. The strongest in-medium
modifications are observed for the nonstrange baryons 
($\Delta m_i^\ast/m_i \approx 15 \%$). Mesons and strange baryons
show smaller changes of the effective masses, e.g. about $10\%$ for
$\Lambda, \pi, K^\ast$, about $5\%$ for the Kaons and nearly no change for the $\Xi$s. 

These results, together with the steep $\chi^2$ contours from
Fig.~2, suggest that the relative particle abundances ``freeze''
shortly \cal{after}  the spontaneous breaking of chiral symmetry.
The success of our fit suggests extremely rapid
chemical equilibration (through abundance-changing reactions)
in the state with broken symmetry. Fig.~2 shows that 
the chemical composition of the hadronic system has to change
substantially within a small temperature interval, just before freeze-out,
even for the crossover transition 
(i.e.\ parameter set CI); for reference, we have
indicated the dynamical path in the $T-\mu_B$ plane corresponding to
the expansion of a perfect fluid (i.e.\ with constant entropy per net
baryon \cite{sub86}).
While $2\to n$ reactions are perhaps too slow to explain such
rapid chemical equilibration~\cite{bass_ch,bass00}, $m\to n$ processes
with several particles in the initial state may be important
as well~\cite{rapp00,grei01,grei02,grei02b}.

Alternatively, the appearence of chemical equilibrium right after the
phase transition (or the crossover) to the state of broken chiral symmetry
might just be the outcome of the dynamical symmetry breaking
process itself~\cite{scav02}, with statistical occupation of the various
hadronic channels according to phase 
space~\cite{beca97,beca98,stock99a,stock99b}.
If so, number-changing reactions in the broken phase need not
proceed at a high rate. To test this picture experimentally,
it might be useful to consider central collisions of small ions like
protons or deuterons, at similar energy and particle densities in the central
region as for central Au+Au. For systems of transverse extent comparable
to the correlation lengths of the chiral condensates,
the dynamical symmetry breaking process should be different from that
in large systems (for example, the mean field approximation should not
apply). The correlation lengths $\xi_{\sigma,\zeta}$ are given by
\begin{equation}
\xi_{\sigma}^{-2} = \frac{\partial^2 \left(\Omega/V\right)}
{\partial \sigma^2}~,
\end{equation}
and accordingly for $\xi_{\zeta}$. We evaluate the curvature of the
thermodynamical potential at the global minimum and for $T$, $\mu_B$,
$\mu_S$ at the freeze-out point. For parameter sets CI, CII, CIII we obtain
$\xi_\sigma=0.37$~fm, 0.41~fm, 0.40~fm, respectively. For the correlation
length of the strange condensate we obtain
$\xi_\zeta=0.20$~fm in all three cases.
The correlation lengths are not very much smaller than, say, the
radius of a proton. Thus, even if the freeze-out point for high-energy
$pp$ collisions happens to be
close to that for Au+Au collisions at RHIC energies,
the transition from the symmetry restored to the broken phase might
be different. Finally, we also note that the correlation lengths obtained
from our effective potential are not larger than the thermal
correlation length $1/T$ at freeze-out, and so corrections beyond the
mean-field approximation employed here should be analyzed in the future.

\section{conclusion}
\label{conclusion}
Particle ratios as calculated in a chiral $SU(3) \,\sigma - \omega$
model are compared with RHIC data for Au+Au at $\sqrt{130} \mbox{ AGeV}$
and with noninteracting gas calculations. 
Since different versions of the chiral model show qualitatively different phase
transition scenarios, we investigate whether the particle
production, i.e. the chemistry of the system, is sensitive to the phase
transition behaviour. Since we have shown that the current data are 
described by all three
different phase transition scenarios and the ideal gas model, we can
so far not favour or rule out any one scenario.
In all interacting models the effective
masses at freeze-out are shifted up to $15\%$ from their vacuum values.
The fitted chemical freeze-out temperatures and chemical   
potentials depend on the order of the phase transition. 
The crossover case yields $15$~MeV shifted $T$ values as compared 
to the noninteracting gas
model while the models with a first order phase transition yield
more than $30$~MeV lower temperatures.
Furthermore, the fitted freeze-out points are located practically right on the
phase transition boundary, in the first order phase transition
scenarios, but $T$ is always $\le T_c$. 
\emph{This suggests that at RHIC the system emerges after the chiral 
chiral phase transition. This of course is only true 
if a first order phase
transition does actually occur in QCD at small chemical potentials 
and high $T$.} ''Freezing'' 
of the relative abundances of various hadrons in the
symmetric phase (at $T > T_c$) is excluded.

\begin{acknowledgements}
The authors are grateful to P. Braun-Munzinger, A. Dumitru, 
L. Gerland, M. Gorenstein, I. Mishustin, and K. Paech
for fruitful discussions. This work is supported by Deutsche 
Forschungsgemeinschaft (DFG), Gesellschaft f\"ur Schwerionenforschung
(GSI), Bundesministerium f\"ur Bildung und Forschung (BMBF), the
Graduiertenkolleg Theoretische und Experimentelle Schwerionenphysik
and by the U.S. Department of Energy, Nuclear Physics
Division (Contract No. W-31-109-Eng-38).
\end{acknowledgements}
\bibliography{chiral}
\bibliographystyle{prsty}
\clearpage
\end{document}

%% file: define.tex
\newcommand{\bea}{\begin{eqnarray}}
\newcommand{\eea}{\end{eqnarray}}
\newcommand{\be}{\begin{equation}}
\newcommand{\ee}{\end{equation}}
\newcommand{\bt}{\begin{tabular}}
\newcommand{\et}{\end{tabular}}

\newcommand{\no}{\nonumber}

\newcommand{\beas}{\begin{eqnarray*}}
\newcommand{\eeas}{\end{eqnarray*}}

%% file: ratios-rhic.bbl
\begin{thebibliography}{10}

\bibitem{ferm50}
E. Fermi, Prog. Theor. Phys. {\bf 5},  570  (1950).

\bibitem{land53}
L.~D. Landau, Izv. Akad. Nauk SSSR Ser. Fiz. {\bf 17},  51  (1953).

\bibitem{hahn86}
D. Hahn and H. St\"ocker, Nucl. Phys. {\bf A452},  723  (1986).

\bibitem{bec00}
F. Becattini, J. Cleymans, A. Keranen, E. Suhonen, and K. Redlich, Phys. Rev.
  {\bf C64},  024901  (2001).

\bibitem{stoe78}
H. St{\"o}cker, W. Greiner, and W. Scheid, Z. Phys. A {\bf 286},  121  (1978).

\bibitem{brau01}
P. Braun-Munzinger, D. Magestro, K. Redlich, and J. Stachel, Phys. Lett. {\bf
  B518},  41  (2001).

\bibitem{west76}
G.~D. Westfall {\it et~al.}, Phys. Rev. Lett. {\bf 37},  1202  (1976).

\bibitem{cser86}
L.~P. Csernai and J.~I. Kapusta, Phys. Rept. {\bf 131},  223  (1986).

\bibitem{brau96}
P. Braun-Munzinger, J. Stachel, J.~P. Wessels, and N. Xu, Phys. Lett. B {\bf
  365},  1  (1996).

\bibitem{raf01}
J. Rafelski, J. Letessier, and G. Torrieri, Phys. Rev. {\bf C64},  054907
  (2001).

\bibitem{Let00}
J. Letessier and J. Rafelski, Int. J. Mod. Phys. {\bf E9},  107  (2000).

\bibitem{brau98}
P. Braun-Munzinger and J. Stachel, Nucl. Phys. {\bf A638},  3  (1998).

\bibitem{cley99}
J. Cleymans and K. Redlich, Phys. Rev. {\bf C60},  054908  (1999).

\bibitem{springer}
D. Zschiesche, P. Papazoglou, S. Schramm, C. Beckmann, J. Schaffner-Bielich, H.
  Stocker, and W. Greiner, Springer Tracts in Modern Physics {\bf 163},  129
  (2000).

\bibitem{paper2}
P. Papazoglou, S. Schramm, J. Schaffner-Bielich, H. St\"ocker, and W. Greiner,
  Phys. Rev. C {\bf 57},  2576  (1998).

\bibitem{paper3}
P. Papazoglou, D. Zschiesche, S. Schramm, J. Schaffner-Bielich, H. St\"ocker,
  and W. Greiner, Phys. Rev. C {\bf 59},  411  (1999).

\bibitem{thei83}
J. Theis, G. Graebner, G. Buchwald, J.~A. Maruhn, W. Greiner, H. Stocker, and
  J. Polonyi, Phys. Rev. D {\bf 28},  2286  (1983).

\bibitem{scha91}
J. Schaffner, I.~N. Mishustin, L.~M. Satarov, H. St{\"o}cker, and W. Greiner,
  Z. Phys. {\bf A341},  47  (1991).

\bibitem{BroRho}
G.~E. Brown and M. Rho, Phys. Rep. {\bf 269},  333  (1996).

\bibitem{mich01}
M. Michalec, W. Florkowski, and W. Broniowski, Phys. Lett. {\bf B520},  213
  (2001).

\bibitem{zsch01}
D. Zschiesche, S. Schramm, H. Stocker, and W. Greiner, Phys. Rev. {\bf C65},
  064902  (2002).

\bibitem{Karsch:2000kv}
F. Karsch, E. Laermann, and A. Peikert, Nucl. Phys. {\bf B605},  579  (2001).

\bibitem{kar98}
F. Karsch, hep-lat/9903031  (1998).

\bibitem{serot97}
B.~D. Serot and J.~D. Walecka, Int. J. Mod. Phys. E {\bf 6},  515  (1997).

\bibitem{sche80}
J. Schechter, Phys. Rev. D {\bf 21},  3393  (1980).

\bibitem{zsch00}
D. Zschiesche {\it et~al.}, Nucl. Phys. {\bf A681},  34  (2001).

\bibitem{bass98}
S.~A. Bass {\it et~al.}, Prog. Part. Nucl. Phys. {\bf 41},  225  (1998).

\bibitem{urqmdhome}
UrQMD Collaboration, http://www.th.physik.uni-frankfurt.de/~urqmd/.

\bibitem{mich_dr}
M. Michalec, Ph.D. thesis, Institute of Nuclear Physics, Crakow, 2001.

\bibitem{flor01}
W. Florkowski, W. Broniowski, and M. Michalec, Acta Phys. Polon. {\bf B33},
  761  (2002).

\bibitem{xu01}
N. Xu and M. Kaneta, Nucl. Phys. {\bf A698},  306  (2002).

\bibitem{zsch02}
D. Zschiesche {\it et~al.} (unpublished).

\bibitem{cley98}
J. Cleymans and K. Redlich, Phys. Rev. Lett. {\bf 81},  5284  (1998).

\bibitem{sub86}
P.~R. Subramanian, H. Stocker, and W. Greiner, Phys. Lett. {\bf B173},  468
  (1986).

\bibitem{bass_ch}
S.~A. Bass, P. Danielewicz, S. Pratt, and A. Dumitru, J. Phys. {\bf G27},  635
  (2001).

\bibitem{bass00}
S.~A. Bass and A. Dumitru, Phys. Rev. {\bf C61},  064909  (2000).

\bibitem{rapp00}
R. Rapp and E.~V. Shuryak, Phys. Rev. Lett. {\bf 86},  2980  (2001).

\bibitem{grei01}
C. Greiner and S. Leupold, J. Phys. {\bf G27},  L95  (2001).

\bibitem{grei02}
C. Greiner, J. Phys. {\bf G28},  1631  (2002).

\bibitem{grei02b}
C. Greiner, Nucl. Phys. {\bf A698},  591  (2002).

\bibitem{scav02}
O. Scavenius, A. Dumitru, and J.~T. Lenaghan,   (2002), hep-ph/0201079.

\bibitem{beca97}
F. Becattini and U.~W. Heinz, Z. Phys. {\bf C76},  269  (1997).

\bibitem{beca98}
F. Becattini, M. Gazdzicki, and J. Sollfrank, Nucl. Phys. {\bf A638},  403
  (1998).

\bibitem{stock99a}
R. Stock, Nucl. Phys. {\bf A661},  282  (1999).

\bibitem{stock99b}
R. Stock, Phys. Lett. {\bf B456},  277  (1999).

\end{thebibliography}
